\documentclass[showpacs,twocolumn,floatfix,superscriptaddress]{revtex4}
\usepackage{graphicx}
 
\newcommand{\bq}{\begin{equation}} 
\newcommand{\ee}{\end{equation}} 
\newcommand{\fr}[2]{\frac{#1}{#2}} 
\newcommand{\eps}{\varepsilon} 
 
\begin{document}
\title{Adiabatic quantization of Andreev levels}
\author{P. G. Silvestrov}
\affiliation{Instituut-Lorentz, Universiteit Leiden, P.O. Box 9506, 2300 RA
Leiden, The Netherlands}
\affiliation{Budker Institute of Nuclear Physics, 630090 Novosibirsk, Russia}
\author{M. C. Goorden}
\affiliation{Instituut-Lorentz, Universiteit Leiden, P.O. Box 9506, 2300 RA
Leiden, The Netherlands}
\author{C. W. J. Beenakker}
\affiliation{Instituut-Lorentz, Universiteit Leiden, P.O. Box 9506, 2300 RA
Leiden, The Netherlands}
\date{9 August 2002}
\begin{abstract}
We identify the time $T$ between Andreev reflections as a classical adiabatic
invariant in a ballistic chaotic cavity (Lyapunov exponent $\lambda$), coupled
to a superconductor by an $N$-mode constriction. Quantization of the
adiabatically invariant torus in phase space gives a discrete set of periods
$T_{n}$, which in turn generate a ladder of excited states
$\varepsilon_{nm}=(m+1/2)\pi\hbar/T_{n}$. The largest quantized period is the
Ehrenfest time $T_{0}=\lambda^{-1}\ln N$. Projection of the invariant torus
onto the coordinate plane shows that the wave functions inside the cavity are
squeezed to a transverse dimension $W/\sqrt{N}$, much below the width $W$ of
the constriction.
\end{abstract}
\maketitle

The notion that quantized energy levels may be associated with classical 
adiabatic
invariants goes back to Ehrenfest and the birth of quantum mechanics
\cite{Ehr16}. It was successful in providing a semiclassical quantization 
scheme
for special integrable dynamical systems, but failed to describe the generic
nonintegrable case. Adiabatic invariants play an interesting but minor role in
the quantization of chaotic systems \cite{Mar89,Gut90}.

Since the existence of an adiabatic invariant is the exception rather than the
rule, the emergence of a new one quite often teaches us something useful about
the system. An example from condensed matter physics is the quantum Hall
effect, in which the semiclassical theory is based on two adiabatic 
invariants:
The flux through a cyclotron orbit and the flux enclosed by the orbit 
center as
it slowly drifts along an equipotential \cite{QHE}. The strong magnetic field
suppresses chaotic dynamics in a smooth potential landscape, rendering the
motion quasi-integrable.

Some time ago it was realized that Andreev reflection has a similar effect on
the chaotic motion in an electron billiard coupled to a superconductor
\cite{Kos95}. An electron trajectory is retraced by the hole that is produced
upon absorption of a Cooper pair by the superconductor. At the Fermi energy
$E_{F}$ the dynamics of the hole is precisely the time reverse of the electron
dynamics, so that the motion is strictly periodic. The period from electron to
hole and back to electron is twice the time $T$ between Andreev reflections.
For finite excitation energy $\varepsilon$ the electron (at energy
$E_{F}+\varepsilon$) and the hole (at energy $E_{F}-\varepsilon$) follow
slightly different trajectories, so the orbit does not quite close and drifts
around in phase space. This drift has been studied in a variety of contexts
\cite{Kos95,Stone96,Shy98,Ada02,Wie02}, but not in connection with adiabatic 
invariants
and the associated quantization conditions. It is the purpose of this paper to
make that connection and point out a striking physical consequence: The wave
functions of Andreev levels fill the cavity in a highly
nonuniform ``squeezed'' way, which has no counterpart in normal state
chaotic or regular billiards. In particular the squeezing is distinct 
from periodic orbit scarring
\cite{Hel84} and entirely different from the random superposition of plane
waves expected for a fully chaotic billiard \cite{Oco87}.

Adiabatic quantization breaks down near the excitation gap, and we will 
argue that random-matrix theory~\cite{Mel96} can be used to quantize the
lowest-lying excitations above the gap. This will lead us to a formula for 
the gap that crosses over from the Thouless energy to the inverse
Ehrenfest time as the number of modes in the point contact is increased.

\begin{figure}
\includegraphics[width=8cm]{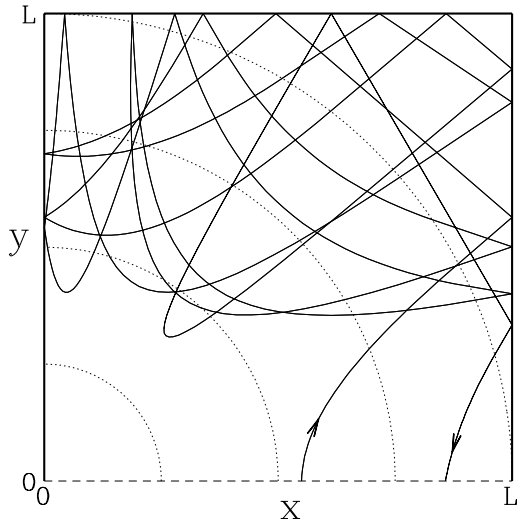}
\caption{
Classical trajectory in an Andreev billiard. Particles in a two-dimensional
electron gas are deflected by the potential $V=[1-(r/L)^{2}]V_{0}$ for $r<L$,
$V=0$ for $r>L$. (The dotted circles are equipotentials.) There is specular
reflection at the boundaries with an insulator (thick solid lines) and Andreev
reflection at the boundary with a superconductor (dashed line). The trajectory
follows the motion between two Andreev reflections of an electron near the
Fermi energy $E_{F}=0.84\,V_{0}$. The Andreev reflected hole retraces this
trajectory in opposite direction.
\label{paden}
}
\end{figure}

\begin{figure}
\includegraphics[width=8cm]{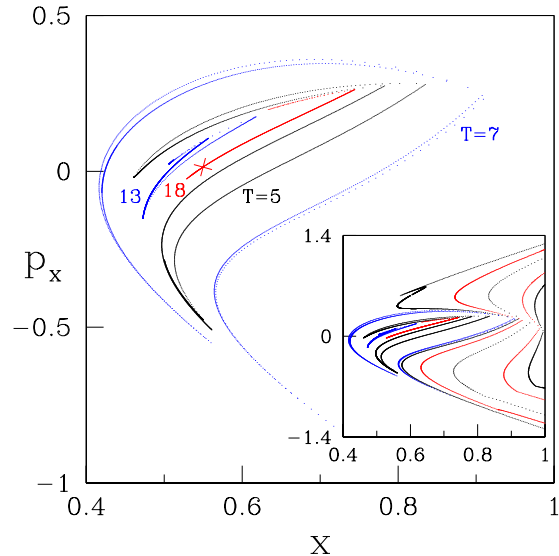}
\caption{
Poincar\'{e} map for the Andreev billiard of Fig.\ \protect\ref{paden}. Each
dot represents a 
starting point of an electron trajectory, 
at position
$x$ (in units of $L$) along the interface $y=0$ and with tangential momentum
$p_{x}$ (in units of $\sqrt{mV_{0}}$). The inset shows the full surface of
section, while the main plot is an enlargement of the central region.  The
drifting quasiperiodic motion follows contours of constant time $T$ between
Andreev reflections. The cross marks the starting point of the trajectory shown
in the previous figure, having $T=18$ (in units of $\sqrt{mL^{2}/V_{0}}$).
\label{phasespace}
}
\end{figure}

To illustrate the problem we represent in Figs.\ \ref{paden} and
\ref{phasespace} the quasiperiodic motion in a particular Andreev billiard. (It
is similar to a Sinai billiard, but has a smooth potential $V$ in the interior
to favor adiabaticity.) Fig.\ \ref{paden} shows a trajectory in real space
while Fig.\ \ref{phasespace} is a section of phase space at the interface with
the superconductor ($y=0$). The tangential component $p_{x}$ of the electron
momentum is plotted as a function of the coordinate $x$ along the interface.
Each point in this Poincar\'{e} map corresponds to one collision of an electron
with the interface. (The collisions of holes are not plotted.) The electron is
retroreflected as a hole with the same $p_{x}$. At $\varepsilon=0$ the
component $p_{y}$ is also the same, and so the hole retraces the path of the
electron (the hole velocity being opposite to its momentum). At non-zero
$\varepsilon$ the retroreflection occurs with a slight change in $p_{y}$,
because of the difference $2\varepsilon$ in the kinetic energy of electrons and
holes. 
The resulting slow drift of the periodic trajectory traces out a contour in the
surface of section. The adiabatic invariant is the function of $x,p_{x}$ that
is constant on the contour. We have found numerically that the drift follows
{\em isochronous\/} contours $C_{T}$ of constant time $T(x,p_{x})$ between
Andreev reflections~\cite{closed}.  
Let us now demonstrate analytically that $T$ is an
adiabatic invariant.

We consider the Poincar\'{e} map $C_{T}\rightarrow C(\varepsilon,T)$ at energy
$\varepsilon$. If $\varepsilon=0$ the Poincar\'{e} map is the identity, so
$C(0,T)=C_{T}$. For adiabatic invariance we need to prove that
$\lim_{\varepsilon\rightarrow 0}dC/d\varepsilon=0$, so that the difference
between $C(\varepsilon,T)$ and $C_{T}$ is of higher order than 
$\varepsilon$~\cite{KAM}.
Since the contour $C(\varepsilon,T)$ can be locally represented by a function
$p_{x}(x,\varepsilon)$, we need to prove that $\lim_{\varepsilon\rightarrow
0}\partial p_{x}(x,\varepsilon)/\partial\varepsilon=0$.

In order to prove this, it is convenient to decompose the map $C_{T}\rightarrow
C(\varepsilon,T)$ into three separate stages, starting out as an electron (from
$C_{T}$ to $C_{+}$), followed by Andreev reflection ($C_{+}\rightarrow C_{-}$),
and then concluded as a hole [from $C_{-}$ to $C(\varepsilon,T)$]. 
Andreev reflection
introduces a discontinuity in $p_{y}$ but leaves $p_{x}$ unchanged, so
$C_{+}=C_{-}$. The flow in phase space as electron ($+$) or hole ($-$) at
energy $\varepsilon$ is described by the action $S_{\pm}({\bf q},\varepsilon)$,
such that ${\bf p}^{\pm}({\bf q},\varepsilon)=\partial S_{\pm}/\partial{\bf q}$
gives the local dependence of (electron or hole) momentum ${\bf
p}=(p_{x},p_{y})$ on position ${\bf q}=(x,y)$. The derivative $\partial
S_{\pm}/\partial\varepsilon=t_{\pm}({\bf q},\varepsilon)$ is the time elapsed
since the previous Andreev reflection. 
Since by construction
$t_{\pm}(x,y=0,\varepsilon=0)=T$ is independent of the position $x$ of the end 
of the trajectory, we find that
$\lim_{\varepsilon\rightarrow 0}\partial
p^{\pm}_{x}(x,y=0,\varepsilon)/\partial\varepsilon=0$, completing the 
proof.

The drift $(\delta x,\delta p_x)$ of a point in the Poincar\'{e} map 
is perpendicular to the vector
$(\partial T/\partial x,\partial T/\partial p_x)$. 
Using also that the map is area preserving, it follows that
\bq
(\delta x,\delta p_x)= \eps f(T)(\partial T/\partial p_x,-\partial T/\partial x)
+O(\eps^2),
\ee
with a prefactor $f(T)$ that is the same along the entire contour.


The adiabatic invariance of isochronous contours may alternatively be obtained
from the adiabatic invariance of the action integral $I$ over the quasiperiodic
motion from electron to hole and back to electron:
\begin{equation}\label{Idef}
I=\oint pdq=\varepsilon\oint\frac{dq}{\dot q}=2\varepsilon T.
\end{equation}
Since $\varepsilon$ is a constant of the motion, adiabatic invariance of $I$
implies adiabatic invariance of the time $T$ between Andreev reflections. This
is the way in which adiabatic invariance is usually proven in text books. Our
proof explicitly takes into account the fact that phase space in the
Andreev billiard consists of two sheets, joined in the constriction
at the interface with the superconductor, with a discontinuity  
in the action on going from one sheet to the other.

The contours of large $T$ enclose a very small area. This will play
a crucial role when we quantize the billiard, so let us estimate the area. It
is convenient for this estimate to measure $p_{x}$ and $x$ in units of the
Fermi momentum $p_{F}$ and width $W$ of the constriction to the
superconductor. The highly elongated shape evident in Fig.\ \ref{phasespace} is
a consequence of the exponential divergence in time of nearby trajectories,
characteristic of chaotic dynamics. 
The rate of divergence is the Lyapunov
exponent $\lambda$. (We consider a fully chaotic phase space.) 
Since the Hamiltonian flow is area
preserving, a stretching $\ell_{+}(t)=\ell_{+}(0)e^{\lambda t}$ of the
dimension in one direction needs to be compensated by a squeezing
$\ell_{-}(t)=\ell_{-}(0)e^{-\lambda t}$ of the dimension in the other
direction. The area $A\simeq\ell_{+}\ell_{-}$ is then time independent.
Initially, $\ell_{\pm}(0)<1$. The constriction at the superconductor acts as a
bottleneck, enforcing $\ell_{\pm}(T)<1$. These two inequalities imply
$\ell_{+}(t)<e^{\lambda(t-T)}$, $\ell_{-}<e^{-\lambda t}$. The enclosed area,
therefore, has upper bound
\begin{equation}
A_{\rm max}\simeq p_{F}We^{-\lambda T}\simeq\hbar Ne^{-\lambda T},\label{Amax}
\end{equation}
where $N\simeq p_{F}W/\hbar\gg 1$ is the number of channels in the point
contact.

We now continue with the quantization. 
The two invariants $\varepsilon$ and $T$
define a two-dimensional torus in the four-dimensional phase space.
Quantization of this adiabatically invariant torus proceeds following
Einstein-Brillouin-Keller \cite{Gut90}, by quantizing the area
\begin{equation}
\oint pdq=2\pi\hbar(m+\nu/4),\;\;m=0,1,2,\ldots\label{EBK}
\end{equation}
enclosed by each of the two topologically independent contours on the torus.
Eq.\ (\ref{EBK}) ensures that the wavefunctions are single valued. (See Ref.\
\cite{Dun02} for a derivation in a two-sheeted phase space.) The integer $\nu$
counts the number of caustics (Maslov index) and in our case should also
include the number of Andreev reflections.

\begin{figure}
\includegraphics[width=8cm]{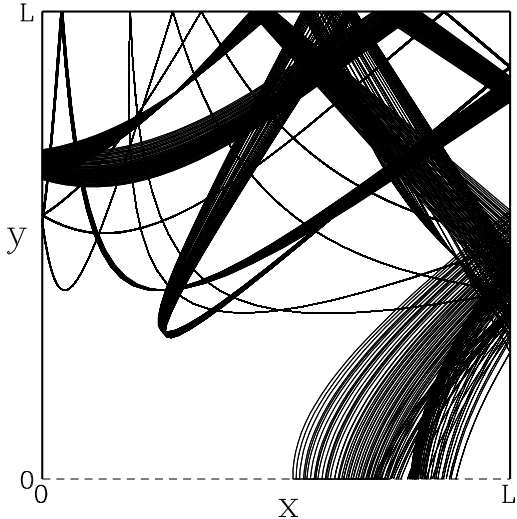}
\caption{
Projection onto the $x$-$y$ plane of the invariant torus with $T=18$,
representing the support of the electron component of the wave function. The
flux tube has a large width near the superconductor, which is squeezed to an
indistinguishably small value after a few collisions with the boundaries.
\label{squeezed}
}
\end{figure}

The first contour follows the quasiperiodic orbit of Eq.\ (\ref{Idef}), leading
to
\begin{equation}
\varepsilon T=(m+{\textstyle\frac{1}{2}})\pi\hbar,
\;\;m=0,1,2,\ldots\label{epsilonTquant}
\end{equation}
The quantization condition (\ref{epsilonTquant}) is
sufficient to determine the smoothed density of states $\rho(\varepsilon)$,
using the classical probability distribution 
$P(T)\propto\exp(-TN\delta/h)$~\cite{Bauer} for
the time between Andreev reflections. (We denote by $\delta$ the level spacing
in the isolated billiard.) The density of states
\begin{equation}
\rho(\varepsilon)=N\int_{0}^{\infty}dT\,P(T)
\sum_{m=0}^{\infty}\delta\bigl(\varepsilon-
(m+{\textstyle\frac{1}{2}})\pi\hbar/T\bigr) \label{rhoBS1}
\end{equation}
has no gap, but vanishes smoothly $\propto\exp(-N\delta/4\varepsilon)$ at
energies below the Thouless energy $N\delta$. This
``Bohr-Sommerfeld approximation'' \cite{Mel96} has been quite successful
\cite{Sch99,Ihr01,Cse02}, but it gives no information on the location of
individual energy levels --- nor can it be used to determine the wave
functions.

To find these we need a second quantization condition, which is provided by the
area $\oint_{T} p_{x}dx$ enclosed by the contours of constant $T(x,p_{x})$,
\begin{equation}
\oint_{T}p_{x}dx=2\pi\hbar(n+\nu/4),\;\;n=0,1,2,\ldots\label{Tquant}
\end{equation}
Eq.\ (\ref{Tquant}) amounts to a quantization of the period $T$, which together
with Eq.\ (\ref{epsilonTquant}) leads to a quantization of $\varepsilon$. For
each $T_{n}$ there is a ladder of Andreev levels
$\varepsilon_{nm}=(m+{\textstyle\frac{1}{2}})\pi\hbar/T_{n}$.

While the classical $T$ can become arbitrarily large, the quantized $T_{n}$ has
a cutoff. The cutoff follows from the maximal area (\ref{Amax}) enclosed by an
isochronous contour. Since Eq.\ (\ref{Tquant}) requires $A_{\rm
max}>2\pi\hbar$, we find that the longest quantized period is
$T_{0}=\lambda^{-1}[\ln N+{\cal O}(1)]$. The lowest Andreev level associated
with an adiabatically invariant torus is therefore
\begin{equation}
\varepsilon_{00}=\frac{\pi\hbar}{2T_{0}}=\frac{\pi\hbar\lambda}{2\ln
N}.\label{epsilon00}
\end{equation}
The time scale $T_{0}\propto|\ln\hbar|$ represents the Ehrenfest time of the
Andreev billiard, which sets the scale for the excitation gap in the
semiclassical limit \cite{Lod98,Tar01,Ada03}.

We now turn from the energy levels to the wave functions. The wave function has
electron and hole components $\psi_{\pm}(x,y)$, corresponding to the two sheets
of phase space. By projecting the invariant torus in a single sheet onto the
$x$-$y$ plane we obtain the support of the electron or hole wave function. This
is shown in Fig.\ \ref{squeezed}, for the same billiard presented in the
previous figures. The curves are streamlines that follow the motion of
individual electrons, all sharing the same time $T$ between Andreev
reflections. (A single one of these trajectories was shown in Fig.\
\ref{paden}.)

Together the streamlines form a flux tube that represents the support of
$\psi_{+}$. The width $\delta W$ of the flux tube is of order $W$ at the 
constriction, but becomes much smaller in the interior of the billiard. Since
$\delta W/W<\ell_{+}+\ell_{-}<e^{\lambda(t-T)}+e^{-\lambda t}$ (with $0<t<T$),
we conclude that the flux tube is squeezed down to a width
\begin{equation}
\delta W_{\rm min}\simeq We^{-\lambda T/2}.\label{deltaW}
\end{equation}
The flux tube for the 
level $\varepsilon_{00}$ has a minimal
width $\delta W_{\rm min}\simeq W/\sqrt{N}$. Particle conservation implies that
$|\psi_{+}|^{2}\propto 1/\delta W$, so that the squeezing of the flux tube is
associated with an increase of the electron density by a factor of $\sqrt{N}$
as one moves away from the constriction.

Let us examine the range of validity of adiabatic quantization.
The drift $\delta x$, $\delta p_{x}$ upon one iteration of the Poincar\'{e} map
should be small compared to $W, p_{F}$. We estimate
\begin{equation}
\frac{\delta x}{W}\simeq\frac{\delta
p_{x}}{p_{F}}
\simeq \fr{\eps_{nm}}{\hbar\lambda N}e^{\lambda T_{n}}
\simeq (m+{\textstyle \fr{1}{2}})\fr{e^{-\lambda (T_0-T_{n})}}{\lambda T_{n}}.
\label{deltaestimate}
\end{equation}
For low-lying levels ($m\sim 1$) the dimensionless drift is $\ll 1$ for $T_n<T_0$.
Even for $T_n=T_0$
one has $\delta x/W\simeq 1/\ln N\ll 1$.

Semiclassical methods allow to quantize only the trajectories with periods
$T\le T_0$. 
The part of phase space with longer periods can be quantized by
random-matrix theory~(RMT), according to which 
the excitation gap $E_{\rm gap}$ is the
inverse of the mean time between Andreev reflections in that part of phase
space \cite{Mel96,Sch99}:
\begin{equation}
{E_{\rm gap}}= \gamma^{5/2}\hbar \fr{\int_{T_{0}}^{\infty}P(T)\,dT}
{\int_{T_{0}}^{\infty}TP(T)\,dT}=
\fr{\gamma^{5/2}\hbar}{T_{0}+2\pi\hbar/N\delta} \ .
\label{crossover}
\end{equation}
Here $\gamma=\fr{1}{2}(\sqrt{5}-1)$ is the golden ratio.
This formula describes the crossover from $E_{\rm gap}=\gamma^{5/2}\hbar/T_0
=\gamma^{5/2}\hbar\lambda/\ln N$
to $E_{\rm gap}=\gamma^{5/2}N\delta/2\pi$ at $N\ln N\simeq\hbar\lambda/\delta$.
It requires $\hbar\lambda /N\delta\gg 1$
(mean dwell time large compared to the Lyapunov time). 
The semiclassical (large-$N$) limit of Eq.~(\ref{crossover}),
$\lim_{N\rightarrow\infty}E_{\rm gap}=0.30\, \hbar/T_0$ is a factor of $5$ 
below the lowest adiabatic level, $\eps_{00}=1.6\, \hbar/T_0$,
so that indeed the energy range near the gap is not accessible by
adiabatic quantization~\cite{last}.

Up to now we considered $2$-dimensional Andreev billiards.
Adiabatic quantization may equally well be applied to $3$-dimensional systems,
with the area enclosed by an isochronous contour as the second adiabatic
invariant. For a fully chaotic phase space with two Lyapunov 
exponents $\lambda_1,\lambda_2$,
the longest quantized period is 
$T_0={\textstyle \fr{1}{2}}(\lambda_1+\lambda_2)^{-1}\ln N$.
We expect interesting quantum size effects on the classical localization
of Andreev levels discovered in Ref.~\onlinecite{Shy98}, which should
be measurable in a thin metal film on a superconducting substrate.

One important challenge for future research is to test the adiabatic
quantization of Andreev levels numerically, by solving the Bogoliubov-De Gennes
equation on a computer. The characteristic signature of the
adiabatic invariant that we have discovered, a narrow region of enhanced
intensity in a chaotic region
that is squeezed as one moves away from the superconductor, should be
readily observable and distinguishable from other features that are unrelated
to the presence of the superconductor, such as scars of unstable periodic
orbits \cite{Hel84}. Experimentally these regions might be observable using a
scanning tunnelling probe, which provides an energy and spatially resolved
measurement of the electron density.

This work was supported by the Dutch Science Foundation NWO/FOM. We thank
\.{I}.~Adagideli and J.~Tworzyd{\l}o for helpful discussions.


\begin{thebibliography}{99}
\bibitem{Ehr16} P. Ehrenfest, Ann.\ Phys.\ (Leipzig) {\bf 51}, 327 (1916).
\bibitem{Mar89} C. C. Martens, R. L. Waterland, and W. P. Reinhardt, J. Chem.\
Phys.\ {\bf 90}, 2328 (1989).
\bibitem{Gut90} M. C. Gutzwiller, {\em Chaos in Classical and Quantum
Mechanics\/} (Springer, Berlin, 1990).
\bibitem{QHE} R. E. Prange, in {\em The Quantum Hall Effect}, edited by R. E.
Prange and S. M. Girvin (Springer, New York, 1990).
\bibitem{Kos95} I. Kosztin, D. L. Maslov, and P. M. Goldbart, Phys.\ Rev.\
Lett.\ {\bf 75}, 1735 (1995).
\bibitem{Stone96} M. Stone, Phys.\ Rev.\ B {\bf 54}, 13222 (1996).
\bibitem{Shy98} A. V. Shytov, P. A. Lee, and L. S. Levitov, Phys.\ Uspekhi {\bf
41}, 207 (1998).
\bibitem{Ada02} \.{I}. Adagideli and P. M. Goldbart, Phys.\ Rev.\ B {\bf 65},
201306 (2002).
\bibitem{Wie02} J. Wiersig, Phys.\ Rev.\ E {\bf 65}, 036221 (2002).
\bibitem{Hel84} E. J. Heller, Phys.\ Rev.\ Lett.\ {\bf 53}, 1515 (1984).
\bibitem{Oco87} P. W. O'Connor, J. Gehlen, and E. Heller, Phys.\ Rev.\ Lett.\
{\bf 58}, 1296 (1987).
\bibitem{Mel96} J. A. Melsen, P. W. Brouwer, K. M. Frahm, and C. W. J.
Beenakker, Europhys.\ Lett.\ {\bf 35}, 7 (1996).
\bibitem{closed} Isochronous contours are defined as $T(x,p_{x})={\rm constant}$ 
at $\eps=0$. We assume that the isochronous contours are 
closed. This is true if the border $p_y=0$ of the classically allowed 
region in the $x,p_x$ section 
is itself an isochronous contour, which is the case if 
$\lim_{y\rightarrow 0}\partial V/\partial y\le0$. 
In this case the particle leaving the superconductor with infinitesimal
$p_y$ can not penetrate into the billiard.
\bibitem{KAM} Adiabatic invariance is defined in the limit
$\eps\rightarrow 0$ and is therefore distinct from invariance in the sense
of Kolmogorov-Arnold-Moser~(KAM), which would require a 
critical $\eps^*$ such
that a contour is exactly invariant for $\eps<\eps^*$. Numerical 
evidence~\cite{Kos95} suggests that the KAM theorem does not apply to a 
chaotic Andreev billiard. 
\bibitem{Dun02} K. P. Duncan and B. L. Gy\"{o}rffy, Ann.\ Phys.\ (New York)
{\bf 298}, 273 (2002).
\bibitem{Bauer} W.~Bauer and G.~F.~Bertsch, Phys. Rev. Lett.,
{\bf 65}, 2213 (1990).
\bibitem{Sch99} H. Schomerus and C. W. J. Beenakker, Phys.\ Rev.\ Lett.\ {\bf
82}, 2951 (1999).
\bibitem{Ihr01} W. Ihra, M. Leadbeater, J. L. Vega, and K. Richter, Europhys.\
J. B {\bf 21}, 425 (2001).
\bibitem{Cse02} J. Cserti, A. Korm\'{a}nyos, Z. Kaufmann, J Koltai, and C. J.
Lambert, Phys.\ Rev.\ Lett.\ {\bf 89}, 057001 (2002).
\bibitem{Lod98} A. Lodder and Yu.\ V. Nazarov, Phys.\ Rev.\ B {\bf 58}, 5783
(1998).
\bibitem{Tar01} D. Taras-Semchuk and A. Altland, Phys.\ Rev.\ B {\bf 64},
014512 (2001).
\bibitem{Ada03} \.{I}. Adagideli and C. W. J. Beenakker, Phys.\ Rev.\ Lett.\ 
{\bf 89}, 237002 (2002).
\bibitem{last} The density of states near the gap is obtained in the same
way as Eq.~(\ref{crossover}), with the result $\rho(\eps)
= c (\eps-E_{\rm gap})^{1/2}N_{\rm eff}^{-1/2}\delta_{\rm eff}^{-3/2}$,
where 
$N_{\rm eff}=N^{1-N\delta/h\lambda}$, 
$\delta_{\rm eff}^{-1}=(\delta^{-1}
+N\ln N/h\lambda)N^{-N\delta/h\lambda}$, and 
$c=4(\pi/\sqrt{5})^{1/2}\gamma^{5/4}
(9+4\sqrt{5})^{2/3}\approx 18$.

\end{thebibliography}
\end{document}